\documentclass[sigconf,screen]{acmart}

\renewcommand\footnotetextcopyrightpermission[1]{} 

\usepackage[utf8]{inputenc}
\usepackage{lmodern}  
\usepackage{amsmath}  
\usepackage{listings}
\usepackage{proof}
\usepackage{hhline}
\usepackage{float}
\usepackage[T1,small]{eulervm}
\usepackage{graphicx}
\usepackage{tikz}
\usepackage{xspace}
\usetikzlibrary{positioning,shapes,arrows}

\usepackage{hyperref}

\newcounter{savefootnote}
\newcounter{symfootnote}
\newcommand{\symfootnote}[1]{%
   \setcounter{savefootnote}{\value{footnote}}%
   \setcounter{footnote}{\value{symfootnote}}%
   \ifnum\value{footnote}>8\setcounter{footnote}{0}\fi%
   \let\oldthefootnote=\thefootnote%
   \renewcommand{\thefootnote}{\fnsymbol{footnote}}%
   \footnote{#1}%
   \let\thefootnote=\oldthefootnote%
   \setcounter{symfootnote}{\value{footnote}}%
   \setcounter{footnote}{\value{savefootnote}}%
}

\newcommand{\sem}[1]{$\llbracket${#1}$\rrbracket$}
\newcommand{\csem}[1]{{\bf #1}}
\newcommand{\var}[1]{{\_}#1}
\newcommand{\lcalc}{$\lambda$-calculus}
\newcommand{\lam}[1]{$\lambda {#1} \cdot$}

\graphicspath{ {./} }

\lstMakeShortInline[columns=fixed]|

\newcommand{\Omit}[1]{}

\AtBeginDocument{%
  \providecommand\BibTeX{{%
    \normalfont B\kern-0.5em{\scshape i\kern-0.25em b}\kern-0.8em\TeX}}}

\makeatletter
\let\@authorsaddresses\@empty
\makeatother

\acmConference[]{}{}{}

\begin{document}

\author{Jayaraj Poroor}
\email{jayaraj.poroor@gmail.com}
\affiliation{%
  \institution{JIFFY.ai}
  \country{}
}

\title[Natural Hoare Logic]
{Natural Hoare Logic: Towards formal verification of programs from logical forms of natural language specifications}

\begin{abstract}
Formal verification provides strong guarantees of correctness of software, which are especially important in safety or security critical systems. Hoare logic is a widely used formalism for rigorous verification of software against specifications in the form of pre-condition/post-condition assertions. The advancement of semantic parsing techniques and higher computational capabilities enable us to extract semantic content from natural language text as formal logical forms, with increasing accuracy and coverage. 

This paper proposes a formal framework for Hoare logic-based formal verification of imperative programs using logical forms generated from compositional semantic parsing of natural language assertions. We call our reasoning approach {\it Natural Hoare Logic}. This enables formal verification of software directly against safety requirements specified by a domain expert in natural language.

We consider both declarative assertions of program invariants and state change as well as imperative assertions that specify commands which alter the program state. We discuss how the reasoning approach can be extended using domain knowledge and a practical approach for guarding against semantic parser errors.
\end{abstract}


\maketitle

\section{Introduction}

Formal verification techniques \cite{woodcock2009formal} are a collection of techniques that can be used to provide strong guarantees of correctness and security for software. They are especially desirable in safety critical software such as the control software of medical devices \cite{li2013improving} or security sensitive portions of a software system \cite{avalle2014formal}. In order to perform formal verification we must specify the correctness or security requirements in a formal language. For instance, consider the following natural language specification of a safety requirement of a medical infusion pump from \cite{murugesan2013compositional}:
\begin{center}
``{\it The infusion manager shall stop infusion whenever a critical alarm occurs.}'' \cite{murugesan2013compositional}
\end{center}

Such natural language specifications are typically given by domain experts who understand the safety requirements of the particular field. However, in order to formally verify the system against these requirements, we must specify them in a formal language, which cannot be typically done by the domain expert. The objective of this paper is to propose a general method that enables software to be formally verified directly against requirements specified by a domain expert in natural language.

Hoare logic \cite{hoare1969axiomatic}\footnote{Also called Floyd-Hoare logic since Floyd had developed a similar system for Flowchart programs earlier\cite{fl001967assigning}.} and its variants are established formal techniques for proving that a program satisfies its specification in the form of pre-condition, post-condition assertions. 

Hoare logic assertions are specified as triples of the form $\{P\} S \{Q\}$, where $P$ and $Q$ are logical assertions on program variables. The meaning of the triple is that: if program variables satisfy the condition $P$ and the program $S$ is executed, then the program variables satisfy the condition $Q$ provided the program terminates. This is called the partial correctness assertion. We may extend it to total correctness with the additional assertion that the program will terminate when starting at a variable state satisfying $P$. In this paper, we consider only partial correctness. If we do not consider exception conditions, for loop-free programs partial correctness coincide with total correctness.

Semantic parsing \cite{kamath2018survey} is a collection of techniques whose objective is to obtain a formal representation of meaning from natural language sentences. For most part, the research in semantic parsing is concerned with obtaining a representation of meaning that can be interpreted by a computer to perform the task at hand, such as querying a database to obtain results. Semantic parsing covers a spectrum of techniques from shallow to deep. Shallow semantic parsing techniques such as semantic role labelling \cite{marquez2008semantic} aims to identify semantic roles (such as agent or location) associated with the events specified in a sentence. In contrast, deep semantic parsing aims to extract a rich representation of meaning known as logical forms in a formal system such as the \lcalc{} \cite{barendregt1984lambda}.

In this paper we concern ourselves with formal logical forms generated by semantic parsers in the form of \lcalc{} expressions. \lcalc{} is preferred  over first order logic for logical forms since it allows us to define semantics compositionally using function composition and function application, from logic forms representing meanings of constituents. Compositional semantics is originally due to Richard Montague \cite{montague1973proper} and hence is broadly called Montague semantics \cite{janssen2011montague}.  Simply-typed $\lambda$-calculus \cite{church1940formulation} with pre-defined constants for logical connectives and quantifiers, known as the higher-order logics\footnote{Higher-order logics may also be considered as a generalization of first order logic.}, is commonly employed as the formal language for logical forms \cite{carpenter1997type}. In this paper we consider logical forms in this formal language\footnote{The lambda calculator available at \url{http://lambdacalculator.com/} is a great educational tool for natural language semantics in the $\lambda$-calculus.}. Owing to the expressive power of $\lambda$-calculus we do not lose any generality. Other logical forms like $\lambda$-DCS \cite{liang2013lambda} can be readily expressed in terms of $\lambda$-calculus.

We propose a framework for associating logical forms generated from assertions about program state made in natural language to pre-condition/post-condition assertions in Hoare logic. We call our framework, {\it Natural Hoare Logic} (NHL). Our objective is to enable formal verification of programs from natural language assertions.

\section{The proposed framework}

We introduce the proposed framework with the help of simple examples. We use the notation \sem{text} to denote the semantics of {\it text} in \lcalc{}. We use \csem{identifier} to denote the semantic function or constant corresponding to {\it identifier} appearing in the sentence. 
 
Consider the statement:
\begin{center}
{\it All balances must be greater than zero.}
\end{center}

where,

\sem{balances} = \lam{x} \csem{balance}(x), a function that returns the balance value, given an account. For the sake of simplicity, in our examples, we restrict our semantic domain (and hence our domain of discourse) to accounts.

\sem{greater than} = \lam{x,y} $x > y$ 

\sem{zero} = \lam{x} \csem{0}, which we may simply write as \csem{0}.

\sem{all} = $\forall$. $\forall$ is simply a higher-order function constant in our logic that takes a truth-valued function {\it f} and evaluates to true if {\it f} evalutes to true for all values.

The semantics of the entire sentence may be composed from the constituent semantics to get:

\sem{All balances must be greater than zero} = 
\begin{equation}
\label{logicform1}
\forall  x \cdot \mathbf{balance}(x) > 0 
\end{equation}

Please note that, $\forall x \cdot P$ is a short-hand for $\forall\ \lambda x \cdot P$. We omit explicit type annotations in the logical forms used in our examples.

Now, consider the following program fragment that simply increments a variable representing balance: 
\begin{equation}
\label{program1}
{\_}\text{balance} := {\_}\text{balance} + 1
\end{equation}
To avoid name confusion we prefix the names of program variables with underscore (e.g., \var{balance}).  

\subsection{Invariant declaratives}

In order to prove that the program holds the invariant specified by the logical form in Equation \ref{logicform1}, we must relate the variables occuring in the logical form to the variables occuring in the program.

We specify the relation between the logical form and program variables via a formal relation. We call this the {\it logical form - program logic} relation (LFPL relation). This is similar in spirit to the state relations \cite{woodcock1996using} from the refinement theory \cite{de1998data} used to prove formal correspondence (forward/backward simulations) between an abstract specification and its concrete implementation. 

We may use the semantic expressions corresponding to the entities present in the natural language specification to construct the relation, as below.

\begin{equation}
\mathcal{R} \triangleq \exists{x} \cdot \mathbf{balance}(x) = \_\text{balance} 
\end{equation}

It may be noted that $\exists{x} \cdot P$ = $\exists\ \lambda{x} \cdot P$.

Given a logical form $\mathcal{L}$ specifying a program invariant\footnote{We consider the invariant to hold true only before and after the program execution, not in the intermediate states.} and an LFPL relation $\mathcal{R}$, we may prove that the program {\it S} satisfies the invariant by establishing the Hoare triple $\{I\}\ S\ \{I\}$, where {\it I} is the strongest condition that satisfies:
\begin{align}
\label{inv}
\mathcal{L} \land \mathcal{R} \implies I \centernot\implies \mathcal{R}
\end{align}

The condition, $I$ $\centernot\implies \mathcal{R}$ is required to remove the correspondance relation from the Hoare logic assertions. In practice, {\it I} should be an assertion on program variables without any correspondence to logical forms. Presence of the correspondence relation in Hoare logic assertions may cause the proof to fail owing to the predicate transformer nature of Hoare logic rules. The Hoare logic rules will transform the program variables in the assertions while the variables in the logical forms remain unchanged. This results in the LFPL relation specifications in the assertions to become inconsistent as Hoare logic rules get applied.

The requirement that {\it I} is the strongest condition that satisfies the above constraint ensures that a weakened form of the user requirement is not take up for proof.

\begin {figure}[t!]
\footnotesize
\begin {tikzpicture}[-latex ,auto ,node distance =1 cm and 4.75cm ,on grid ,
semithick]
\node (P) {\{I\}};
\node (S) [below of=P] {S};
\node (Q) [below=of S] {\{I\}};
\node (L) [left=of S] {$\mathcal{L}$};
\path (L) edge node[above] {$\mathcal{R}$} (P);
\path (L) edge node[above] {$\mathcal{R}$} (Q);
\path (S) edge node[left] {} (P);
\path (Q) edge node[left] {} (S);

\end{tikzpicture}
\caption{Invariant proofs in Hoare logic from logical forms.}
\label{fig:invproof}
\end{figure}
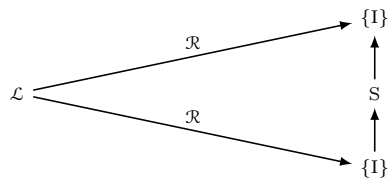

Figure \ref{fig:invproof} visualizes how $\mathcal{R}$ relates the invariant specified in the logical form $\mathcal{L}$ to pre-condition $P$ and post-condition $Q$. The arrow $\{Q\} \rightarrow S \rightarrow \{P\}$ shows the direction of the proof. Program execution and the state transformation occurs in the opposite direction.
 
Using the Hoare triple in Equation \ref{inv} for proving logical form \ref{logicform1} on program \ref{program1}:

Since:
\begin{align}
\{\forall  x \cdot \mathbf{balance}(x) > 0 \} \land \exists{x} \cdot \mathbf{balance}(x) = \_\text{balance}\\
\cline{1-1}
\_\text{balance} > 0
\end{align}

We need to prove:
\begin{align}
\{\_\text{balance} > 0\}\\
{\_}\text{balance} := {\_}\text{balance} + 1\\
\{\_\text{balance} > 0\}
\end{align}

We can prove this by considering a stronger post-condition {\it \var{balance} $> 1$} and applying the assignment rule, and then applying the consequence rule to weaken the post-condition to get back the invariant.

The LFPL relation simply specifies the formal association between the logical form and the program variables. We may make a general assertion in natural language and then use the LFPL relation to establish the association. For example, consider the natural language assertion:

\begin{center}
{\it All values must be greater than zero.}
\end{center}

Its corresponding logical form may be:
\begin{equation}
\forall x \cdot \mathbf{valueof}(x) > 0
\end{equation}

In this case, the LFPL relation would specify that the program variable \_{balance} is the value of an account, as follows:

\begin{equation}
\exists x \cdot \mathbf{valueof}(x) = {\_}\text{balance}
\end{equation}

\subsection{Imperatives}

Consider the statement:
\begin{center}
{\it Increment the balance.}
\end{center}

This is an imperative sentence, specifying a command to be performed. Semantics of imperatives are much more challenging than that of declarative statements \cite{ kaufmann2009unified}. A common approach for specifying semantics of imperative sentences is by introducing modal operators\cite{oikonomou2016imperatives, kaufmann2019fine}. In this paper, we do not consider imperatives in a general setting, instead restrict ourselves to considering only those imperative statements that specify operations on program state.

As a result, we explicitly model a restricted form of modality as a logical relation between the current state of the world (before-state) and the state after the command has been carried out (after-state)\footnote{We may consider this as a modality with two possible worlds represented by the before-state and after-state.}. We define a function {\it post} that maps elements to their after-state. Therefore, $\lambda x \cdot balance(post(x))$ gives the balance of an account after the specified command has been carried out. The semantics of `increment' can be defined as a higher order function as follows:

\sem{increment} = $\lambda f \cdot \forall x \cdot f(post(x)) = f(x) + 1$ 

This allows such imperative commands to be composed with the semantics of its objects, enabling compositional semantics, e.g, 

\sem{increment the balance} = $\forall x \cdot \mathbf{balance}(post(x)) = \mathbf{balance}(x) + 1$

Of course we are unable to model general imperative statements such as: {\it you may increment the balance}. However, we believe our approach is useful enough for specifying pre-condition/post-conditions assertions on sequential imperative programs.

In addition to the $\mathcal{R}$ relation defined previously, we also define an $\mathcal{R}'$ relation as:

\begin{equation}
\mathcal{R'} \triangleq \exists{x} \cdot \mathbf{balance}(post(x)) = \_\text{balance} 
\end{equation}

$\mathcal{R}'$ is obtained by replacing the variable $x$ by $post(x)$ in the body of $\mathcal{R}$, assuming the $\mathcal{R}$ relation is on a single logical variable $x$. If there are multiple logical variables, then the $post$ function must be applied on all of them. 

Given a $\mathcal{L}$ which is a logical form of the imperative assertion, we may prove that a program $S$ satisfies the specified imperative command, by proving the following:

\begin{align}
\label{inv}
\{\mathcal{L} \land \mathcal{R}\}\ S\ \{\mathcal{L} \land \mathcal{R}'\}\\
\end{align}

This is simpler than that of the invariant because post-condition and pre-condition use different LFPL relations, $\mathcal{R}$ and $\mathcal{R}'$.

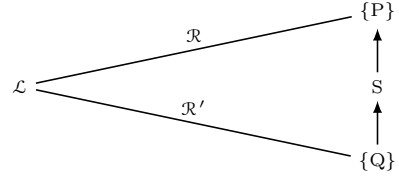
\begin {figure}[t!]
\footnotesize
\begin {tikzpicture}[-latex ,auto ,node distance =1 cm and 4.75cm ,on grid ,
semithick]
\node (P) {\{P\}};
\node (S) [below of=P] {S};
\node (Q) [below=of S] {\{Q\}};
\node (L) [left=of S] {$\mathcal{L}$};
\path (L) edge[-] node[above] {$\mathcal{R}$} (P);
\path (L) edge[-] node[above] {$\mathcal{R}'$} (Q);
\path (S) edge node[left] {} (P);
\path (Q) edge node[left] {} (S);

\end{tikzpicture}
\caption{Hoare logic proofs from imperative logical forms.}
\label{fig:impproof}
\end{figure}

Figure \ref{fig:impproof} visualizes how LFPL relations $\mathcal{R}$ and $\mathcal{R}'$ are used for Hoare logic proofs from imperative logical forms.

For the sentence, {\it increment the balance} the Hoare triple to prove becomes:

We need to prove:
\begin{align}
\{ \mathcal{L} \land \exists{x} \cdot \mathbf{balance}(x) = \_\text{balance} \}\\
{\_}\text{balance} := {\_}\text{balance} + 1\\
\{ \mathcal{L} \land \exists{x} \cdot \mathbf{balance}(post(x)) = \_\text{balance} \}
\end{align}
where $\mathcal{L} = \forall x \cdot \mathbf{balance}(post(x)) = \mathbf{balance}(x) + 1$.

Applying the assignment rule to the post-condition, we get the following assertion, from which the pre-condition follows:

\begin{equation}
\mathcal{L} \land \exists{x} \cdot \mathbf{balance}(post(x)) = \_\text{balance} + 1
\end{equation}

\subsection{Conditional imperatives}

We consider conditional imperatives where the conditional and imperative parts are specified separately. An example of a conditional imperative statement is:

\begin{center}
IF: {\it the balance is greater than 0} THEN: {\it increment the balance.}
\end{center}

The proof approach for imperatives may be extended to conditional imperatives in a straightforward manner. 

If $\mathcal{E}$ is the logical form of the conditional part and $\mathcal{L}$, then the proof rule is:
\begin{align}
\label{inv}
\{\mathcal{L} \land \mathcal{R} \land \mathcal{E}\}\ S\ \{\mathcal{L} \land \mathcal{R}'\}\\
\end{align}

\begin {figure}[t!]
\footnotesize
\begin {tikzpicture}[-latex ,auto ,node distance =1 cm and 4.75cm ,on grid ,
semithick]
\node (P) {\{P\}};
\node (S) [below of=P] {S};
\node (Q) [below=of S] {\{Q\}};
\node (L) [left=of P] {$\mathcal{L} \land \mathcal{E}$};
\node (L1) [below=of L] {$\mathcal{L}$};
\path (L) edge[-] node[above] {$\mathcal{R}$} (P);
\path (L1) edge[-] node[above] {$\mathcal{R}'$} (Q);
\path (L) edge[-] node[above] {} (L1);
\path (S) edge node[left] {} (P);
\path (Q) edge node[left] {} (S);

\end{tikzpicture}
\caption{Hoare logic proofs from logical forms of conditional imperatives.}
\label{fig:condimp}
\end{figure}
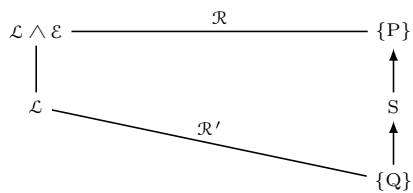

The proof rule is visually illustrated in Figure \ref{fig:condimp}.

\subsection{Pre/Post-condition declaratives}

Another form of natural language specification we consider is the declarative specification of the pre-condition and post-condition separately. For instance:

\begin{center}
IF: {\it the balance is greater than 0} THEN AFTER: {\it balance must be remain non-negative}
\end{center}

This case is simpler than the imperative specification. The Hoare logic assertion would be:

\begin{center}
\begin{align}
\label{inv}
\{\mathcal{L} \land \mathcal{R}\}\ S\ \{\mathcal{L'} \land \mathcal{R}'\}\\
\end{align}
\end{center}

where $\mathcal{L'}$ is the logical form of the post-condition and $\mathcal{L}$ is the logical form of the pre-condition.

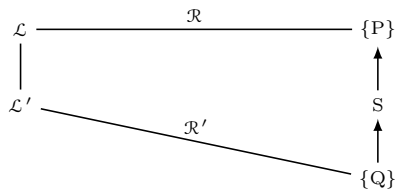
\begin {figure}[t!]
\footnotesize
\begin {tikzpicture}[-latex ,auto ,node distance =1 cm and 4.75cm ,on grid ,
semithick]
\node (P) {\{P\}};
\node (S) [below of=P] {S};
\node (Q) [below=of S] {\{Q\}};
\node (L) [left=of P] {$\mathcal{L}$};
\node (L1) [below=of L] {$\mathcal{L}'$};
\path (L) edge[-] node[above] {$\mathcal{R}$} (P);
\path (L1) edge[-] node[above] {$\mathcal{R}'$} (Q);
\path (L) edge[-] node[above] {} (L1);
\path (S) edge node[left] {} (P);
\path (Q) edge node[left] {} (S);

\end{tikzpicture}
\caption{Hoare logic proofs from pre/post-condition logical forms.}
\label{fig:prepost}
\end{figure}

Figure \ref{fig:prepost} visualizes how LFPL relations $\mathcal{R}$ and $\mathcal{R}'$ are used for Hoare logic froms for logical forms of pre/post-condition assertions.

\subsection{Extending with domain knowledge}

In this section we discuss how our reasoning technique can be extended using domain knowledge specified in the form of logical expressions. For example, consider the sentence and its logical form below:

\begin{center}
{\it All values must be greater than zero}
\end{center}

Logical form:

\begin{equation}
\forall x \cdot \mathbf{valueof}(x) > 0
\end{equation}

Using the LFPL relation below, assume that we have already established proof for this invariant for some program {\it S}.

\begin{equation}
\exists x \cdot \mathbf{valueof}(x) = \_\text{balance}
\end{equation}

Now consider a new sentence and its logical form:
\begin{center}
{\it All balances must be greater than zero}
\end{center}

\begin{equation}
\label{b1}
\forall x \cdot \mathbf{balance}(x) > 0
\end{equation}

Consider the domain knowledge that the balance of an account is its value, which we may express logically as:
\begin{equation}
\forall x \cdot \mathbf{balance}(x) = \mathbf{valueof}(x)
\end{equation}

We may use this and our previous proof to now infer, that program {\it S} satisfies new invariant in Equation \ref{b1}, without having to go through the proof again.

In many domains, formalized domain knowledge exists in the form of domain ontologies in languages such as the web ontology language\cite{mcguinness2004owl} (which has a logical basis in description logics \cite{baader2004description}). For instance the ontological assertion that {\it A} {\bf isa} {\it B} can be expressed as:
\begin{equation}
\forall x \cdot A(x) \implies B(x)
\end{equation} 

Bringing such knowledge to aid in our reasoning is likely to be quite beneficial.

\subsection{Guarding against semantic parser errors}

Owing to the variety and ambiguity in natural language sentences, semantic parsers are probabilistic parsers. The parse (and the resultant logical form) with the highest probability is taken as the selected parse (logical form). This need not always be the correct parse. Since we are targeting formal verification of programs, we need to guard against these errors. In this section, we consider one practical approach. 

Given a language of logical forms generated by a semantic parser, we define a function from logical forms to natural language, let's call this function $\mathcal{N}$. We may use this function to paraphrase the user input in the manner understood by the system. Given a natural language sentence {\it S} input by a user and if its logical form generated by a semantic parser is $\llbracket{S}\rrbracket$, then we could ask back the user:

\begin{center}
Did you mean: $\mathcal{N}(\llbracket{S}\rrbracket)$?
\end{center}

If the answer is {\it yes} we proceed, otherwise we could show the $\mathcal{N}$-mappings of other probable logical forms (if available) for user to choose from. User's selection may also be used as a form of labelling for retraining the model. If the user is unable to choose, then the system may prompt the user to rephrase the specification (perhaps as simpler sentences).

Of course it is important for the $\mathcal{N}$-function to generate a consistent and understandable natural language representation. This is usually a simpler problem than the semantic parsing itself. 

Semantic parsing based on synchronous context free grammars (SCFG) \cite{wong2007learning} is especially interesting to address this problem. This approach recasts semantic parsing as a machine translation problem - that of translating from natural language to the language of logical forms. This enables the parsing to be inverted to generate natural language statements from logical forms, automically giving us the function $\mathcal{N}$ from the learnt grammar. 

\section{Brief review of related fields}

In this section, we briefly review Hoare logics and semantic parsing since our framework attempts to connect logical forms that are outcomes of semantic parsing with Hoare logic assertions on programs.

\subsection*{Hoare logic}

We briefly review the basic Hoare logic rules. The simplest rule is the axiom of assignment:

\begin{equation}
\infer{\{Q[E/x]\}\  x := E\  \{Q\}}{}
\end{equation}

The notation $Q[E/x]$ stands for the expression obtained after all instances of the variable $x$ occuring in $Q$ is substituted with expression $E$.

The rule for sequential composition can be used to infer the pre-condition/post-condition asserts of a program from its sequential components:

\begin{equation}
\infer{\{P\}\ S_1;S_2\ \{R\} }{\{P\}\ S_1\ \{Q\} \ \ \  \{Q\}\ S_2\ \{R\} }
\end{equation}

The conditional rule may be applied to reason about if-else statements.

\begin{equation}
\infer{\{P\}\ if\ E\ then\ S_1\ else\ S_2 \ \{Q\} }
{\{E \land P\}\ S_1\ \{Q\} \ \ \ \  \{\lnot E \land P\}\  S_2\  \{Q\} }
\end{equation}

The consequence rules allows us to strengthen the pre-condition and weaken the post-condition:

\begin{equation}
\infer{ \{P\}\ S\ \{Q\} }{P{\implies}P' \ \  \ \{P'\}\ S\ \{Q'\} \ \ \ Q'{\implies}Q}
\end{equation}

Reasoning about loops is done using the {\it while} loop rule:

\begin{equation}
\infer{\{I\}\ while\ E\ do\ S\ \{\lnot E \land I\}}
{\{I \land E\}\ S\ \{I\}}
\end{equation}

Here {\it I} is the {\it loop invariant}. One of the primary difficulties in automating Hoare logic proofs is the construction of loop invariants. In general, the code must be manually annotated with loop invariants for Hoare logic proofs to go through. 

The Hoare logic axioms and inference rules provide an axiomatic semantics for an imperative programming language. Hoare logic rules may be recast as rules for obtaining weakest pre-conditions from post-condition assertions (or equivalently, strongest post-conditions from pre-condition assertions). The weakest pre-condition or predicate transformer semantics is originally due to Dijskstra \cite{dijkstra1975guarded}.

In recent years, the basic Hoare logic \cite{apt2019fifty} has been extended to support formal verification of complex and real-world code. For instance, separation logic \cite{reynolds2002separation} is one of its variants used for proving correctness properties of programs with pointer data structures. Crash Hoare Logic \cite{chen2015using} extends Hoare logic with special {\it crash} assertions and recovery procedures that enables formal reasoning about crash recovery. The paper establishes formal proofs using this extended logic for FSCQ file system.

A number of researchers have worked on defining Hoare logics for real-world programming languages, both high level and low level. A Hoare Logic for  Java embedded in the Isabelle/HOL theorem prover\cite{nipkow2002isabelle} is given in \cite{von2001hoare}.  A Hoare logic for realistic machine code is presented in \cite{myreen2007hoare}.

The principle of forward (backward) simulations from the refinement theory \cite{de1998data} may be used to prove formal correspondence between an abstract specification and concrete implementation. This enables us to prove formal properties on the abstract specification using Hoare logic (or equivalent methods) which then carries over to the concrete implementation owing to the simulation proofs. Formal verification of complex programs such as the seL4 microkernel takes this approach \cite{klein2009sel4}.

Mechanization of Hoare logic proofs involve first generating verification conditions by a verification condition generator (VCG) that apply Hoare logic rules on the annotated code \cite{matthews2006verification}. The verification conditions are then typically mechanically checked by an automated theorem prover.

\subsection*{Semantic parsing}

Combinatory categorical grammars (CCGs) \cite{steedman2011combinatory} is a well studied and powerful method to achieve compositional semantics. CCGs employ a few general purpose combinatory rules along with a highly lexicalized grammar that maps tokens/phrases to categories. CCG2Lambda \cite{martinez2016ccg2lambda} is a recent work that compositionally maps CCG trees to $\lambda$-terms.

Compositional semantics can also be achieved by attaching logical forms to context-free syntactic parsing rules (usually parsing is done on tokens/phrases tagged with semantic annotations)\footnote{e.g., SippyCup semantic parser available at \url{https://github.com/wcmac/sippycup}}. Whenever a context-free syntactic rule is applied, the attached logical form is composed with the logical forms of the constituents  to generate a resultant logical form.

Semantic parsing based on synchronous context free grammars (SCFG) is proposed in \cite{wong2007learning}. This approach recasts semantic parsing as a machine translation problem - that of translating from natural language to the language of logical forms. This also enables the parsing to be inverted to generate natural language statements from logical forms. The ability to invert logical forms to natural language statements is especially interesting for the problem being addressed in this paper. This enables us to give a natural language feedback of what the system has understood regarding the natural language specification made by the domain expert. This acts as a check against any potential errors introduced by the semantic parser.

Compositional semantics using  dependency-based parse structures is discussed in \cite{liang2013learning}. More recently, in \cite{reddy2017universal}, a compositional semantics is defined over universal dependency parse trees. Here the $\lambda$-calculus logical forms capture semantics in Neo-Davidsonian or event style \cite{champollion2015interaction}.

Since natural language parsing involves ambiguity, a learnt scoring model (usually a linear or a log-linear model) is applied during the semantic parsing to keep track of most probable parses \cite{liang2015bringing}. Learning weights in the scoring model is done in a supervised setting. Recently techniques for weak supervision wherein the sentences are labelled with the denotations (i.e., the results of executing the logical forms) are being investigated \cite{berant2013semantic, liang2016learning, wang2019learning} replacing the need for strong supervision using parse tree annotations or even annotating sentences with the logical forms. This brings in the problem of program induction, i.e., inducing logical forms from denotations, which is a key challenge that must be solved when labelling with denotations.

Another approach to semantic parsing \cite{shi2005putting} takes advantage of semantic lexicons based on specific linguistic theories such as VerbNet\cite{schuler2005verbnet} which  is based on Levin's theory of verb alternation classes\cite{levin1993english} and FrameNet\cite{baker1998berkeley} which is based on frame semantics \cite{fillmore2006frame}.

A yet another approach is to target a narrow domain specific language such as the structured query language (SQL). In such a case the semantic parser may directly generate the SQL (DSL) or may first generate an intermediate logical form, which is then automatically translated to the DSL (SQL). Natural language to SQL is a well studied problem \cite{zhong2017seq2sql}. The current state of the art supports SQL queries without joins using deep learning models.  Such deep learning models have also become state-of-the-art in semantic role labelling. 

A semantic parser for parsing natural language sentences to if-this-then-that recipes is given in \cite{quirk2015language}, targeting the IFTTT service \cite{ovadia2014automate}. Their natural language specifications are somewhat similar to the conditional imperative forms discussed in this paper.  However, their semantic parser output is in the form of IFTTT recipes rather than $\lambda$-calculus logical forms. Their parsing approach is based on KRISP \cite{kate2007semi} that uses a classifier with a string subsequence kernel \cite{lodhi2002text} for mapping the natural language sentences to most probable production rules. 

\section{Conclusion and Future work}

Hoare logic belongs to a general class of formal systems used for reasoning about programs. The results presented here may also be employed to reason with other program logics such as the dynamic logic\cite{harel2001dynamic}. We intend to further validate the NHL framework with more elaborate examples and case studies. Exploring the use of domain knowledge to aid reasoning, investigating approaches for a formal soundess proof, and mechanization of the framework are other lines of future work.

\bibliographystyle{ACM-Reference-Format}

\bibliography{nhl}

\end{document}